\documentclass[%
 aip,
 amsmath,amssymb,
 reprint,%
]{revtex4-1}

\usepackage{graphicx}
\usepackage{dcolumn}
\usepackage{bm}

\usepackage[utf8]{inputenc}
\usepackage[T1]{fontenc}
\usepackage{mathptmx}
\usepackage[english]{babel} 
\usepackage{hyperref}
\newcommand{\heoxone}{(MgCoNiCuZn)O}
\usepackage{lineno}
\usepackage{xcolor}
\graphicspath{{./figures/}}

\begin{document}

\preprint{AIP/123-QED}

\title{Long-range magnetic ordering in rocksalt-type high-entropy oxides}

\author{Marco Polo Jimenez-Segura}
 \affiliation{Max Planck Institute for Solid State Research, Heisenbergstrasse 1, 70569, Stuttgart, Germany\looseness=-1}
\affiliation{Institut fuer Funktionelle Materie und Quantentechnologie 3, Universitaet Stuttgart, Pfaffenwaldring 57,70569 Stuttgart, Germany\looseness=-1}%

\author{Tomohiro Takayama}%
 \affiliation{Max Planck Institute for Solid State Research, Heisenbergstrasse 1, 70569, Stuttgart, Germany\looseness=-1}
\affiliation{Institut fuer Funktionelle Materie und Quantentechnologie 3, Universitaet Stuttgart, Pfaffenwaldring 57,70569 Stuttgart, Germany\looseness=-1}%

\author{David Bérardan}%
\affiliation{ICMMO (UMR 8182 CNRS), Université Paris-Sud, Université Paris-Saclay, 91405, Orsay, France\looseness=-1}%

\author{Andreas Hoser}%
\affiliation{%
	Helmholtz-Zentrum Berlin für Materialien und Energie, D-14109 Berlin, Germany}%

\author{Manfred Reehuis }%
\affiliation{%
	Helmholtz-Zentrum Berlin für Materialien und Energie, D-14109 Berlin, Germany}%

\author{Hidenori Takagi}
\affiliation{Max Planck Institute for Solid State Research, Heisenbergstrasse 1, 70569, Stuttgart, Germany\looseness=-1}
\affiliation{Institut fuer Funktionelle Materie und Quantentechnologie 3, Universitaet Stuttgart, Pfaffenwaldring 57,70569 Stuttgart, Germany\looseness=-1}%
\affiliation{Department of Physics, University of Tokyo, Hongo 7-3-1, Bunkyo-ku, Tokyo 113-0033}

\author{Nita Dragoe}%
\affiliation{ICMMO (UMR 8182 CNRS), Université Paris-Sud, Université Paris-Saclay, 91405, Orsay, France\looseness=-1}%

\date{\today}

\begin{abstract}
We report the magnetic properties of Mg$_{0.2}$Co$_{0.2}$Ni$_{0.2}$Cu$_{0.2}$Zn$_{0.2}$O, a high-entropy oxide with rocksalt structure, and the influence of substitutions on these properties. From the magnetic susceptibility and neutron diffraction measurements, we found that this compound exhibits long-range magnetic order below 120~K despite the substantial structural disorder. The other rocksalt-type high-entropy oxides with various chemical substitutions were found to host either an antiferromagnetic order or spin-glass state depending on the amount of magnetic ions. The presence of magnetic order for such a disordered material potentially provide a route to explore novel magnetic properties and functions.

\end{abstract}

\maketitle

Akin to the high entropy alloys known for over a decade  \cite{yeh_nanostructured_2004}, stabilization of new phases associated with configurational entropy was shown to manifest in oxides with rocksalt structure  \cite{rost_entropy-stabilized_2015}. The first entropy-stabilized oxide Mg$_{0.2}$Co$_{0.2}$Ni$_{0.2}$Cu$_{0.2}$Zn$_{0.2}$O [denoted as (MgCoNiCuZn)O hereafter] was obtained by quenching the mixture of MgO, NiO, CuO, CoO and ZnO from high temperatures  \cite{rost_entropy-stabilized_2015}. The compound showed a reversible phase transformation between the single rocksalt phase and multiphase state depending on temperature, indicating the critical role of entropy in the formation of a face-centered-cubic (FCC) lattice of randomly distributed 5 cations.

It is likely that similar stabilization can be realized in other classes of compounds. Indeed, the syntheses of entropy-stabilized oxides with perovskite  \cite{jiang_new_2018} and fluorite structures  \cite{chen_five-component_2018} have been achieved recently. Besides oxides, entropy-stabilization has been applied to other ceramics such as borides  \cite{gild_high-entropy_2016} and carbides  \cite{zhou_high-entropy_2018}. In addition, pioneering efforts for developing other synthetic methods of these compounds have already been reported  \cite{kotsonis_epitaxial_2018, biesuz_synthesis_2018, sarkar_nanocrystalline_2017}. We wish to underline here that a reversible phase transformation, similar to the one shown by Rost \textit{et al.} \cite{rost_entropy-stabilized_2015}, should be observed if the materials are actually stabilized by entropy, contrary to formation of a simple multi-component solid solution.

Shortly after the initial report of entropy-stabilized oxides, which we call high-entropy oxides (HEOx), we showed that chemical substitutions onto HEOx is possible, for example by replacing one of the divalent cations with 1:1 mixture of monovalent and trivalent ions, which has greatly increased the compositional space  \cite{berardan_colossal_2016}. These compounds are not only intriguing in terms of physical or solid-state chemistry, but also appeared to possess novel functions. For example, the rocksalt-type HEOx were shown to display colossal dielectric constant  \cite{berardan_colossal_2016} and, in the case of alkali-doped compounds, high ionic conductivity \cite{berardan_room_2016}. They have been also turned out to have high-capacity for Li-ion storage \cite{sarkar_high_2018, qiu_high_2019,qiu_high_2019}.

So far, the magnetic properties of HEOx have been barely investigated \cite{meisenheimer_giant_2017} despite the presence of magnetic ions such as Co$^{2+}$ and Ni$^{2+}$. The magnetic interactions in HEOx are potentially quite complex because of the random distribution of magnetic ions with different electron configurations, and, at a glance, they are expected to display a spin-glass-like state as a consequence of strong structural disorder. As opposed to such a naïve view, we disclose in this paper that the rocksalt-type prototypical HEOx, (MgCoNiCuZn)O, exhibits long-range magnetic ordering in spite of the strong disorder of cations. Additionally, magnetic ground states of rocksalt-type HEOx with various compositions were found to be either long-range magnetic order or spin-glass depending on the content of magnetic ions. The magnetic properties of HEOx can thus be finely tuned by varying their composition.

\begin{figure}[t]
	\includegraphics[width=8cm]{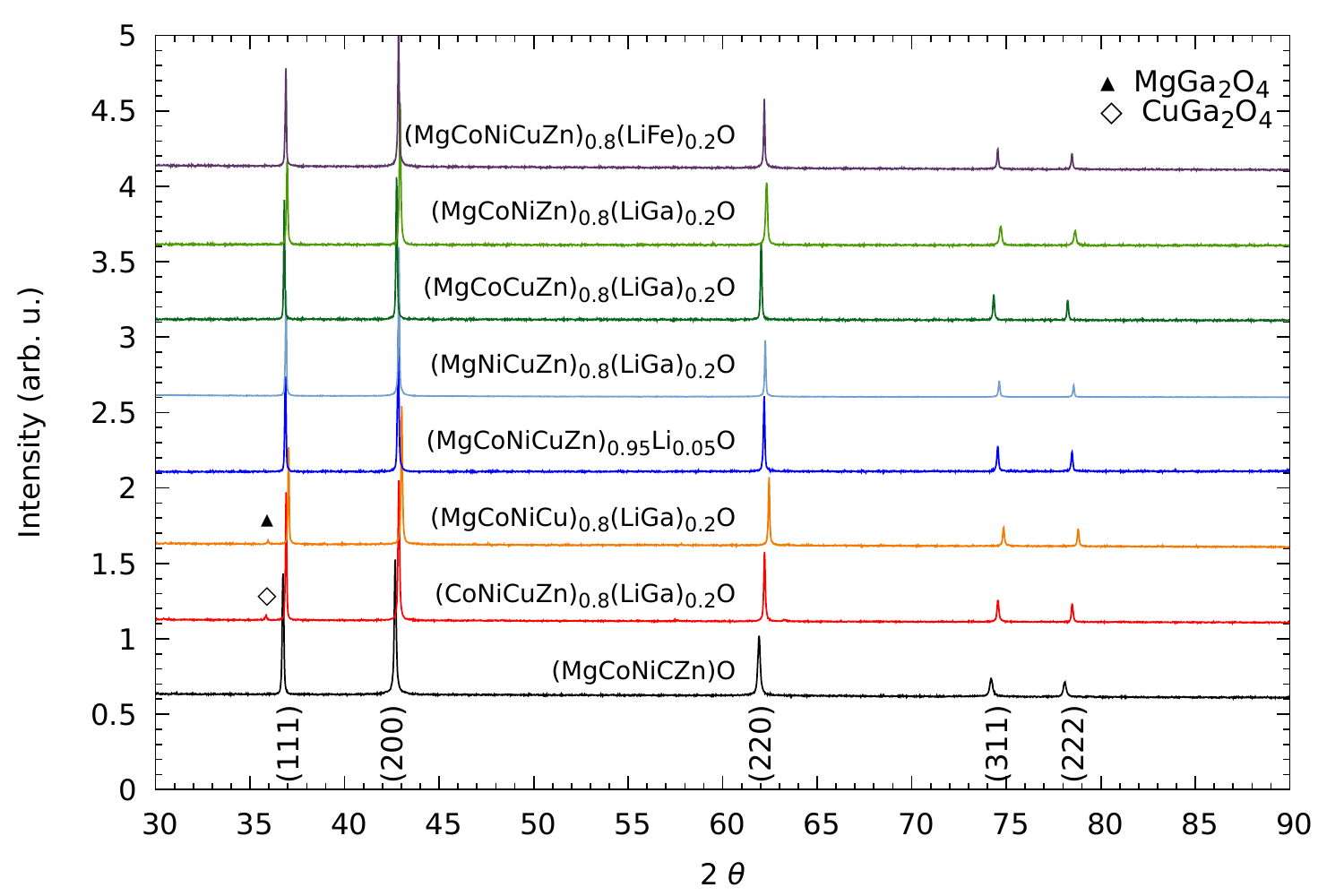}
		\setlength{\abovecaptionskip}{5pt}
		\setlength{\belowcaptionskip}{-14pt}
		\caption{\label{fig:XRD} (color online) X-ray-diffraction patterns of several high entropy oxides. }
\end{figure}

\begin{figure}[t]
	\includegraphics[width=8cm]{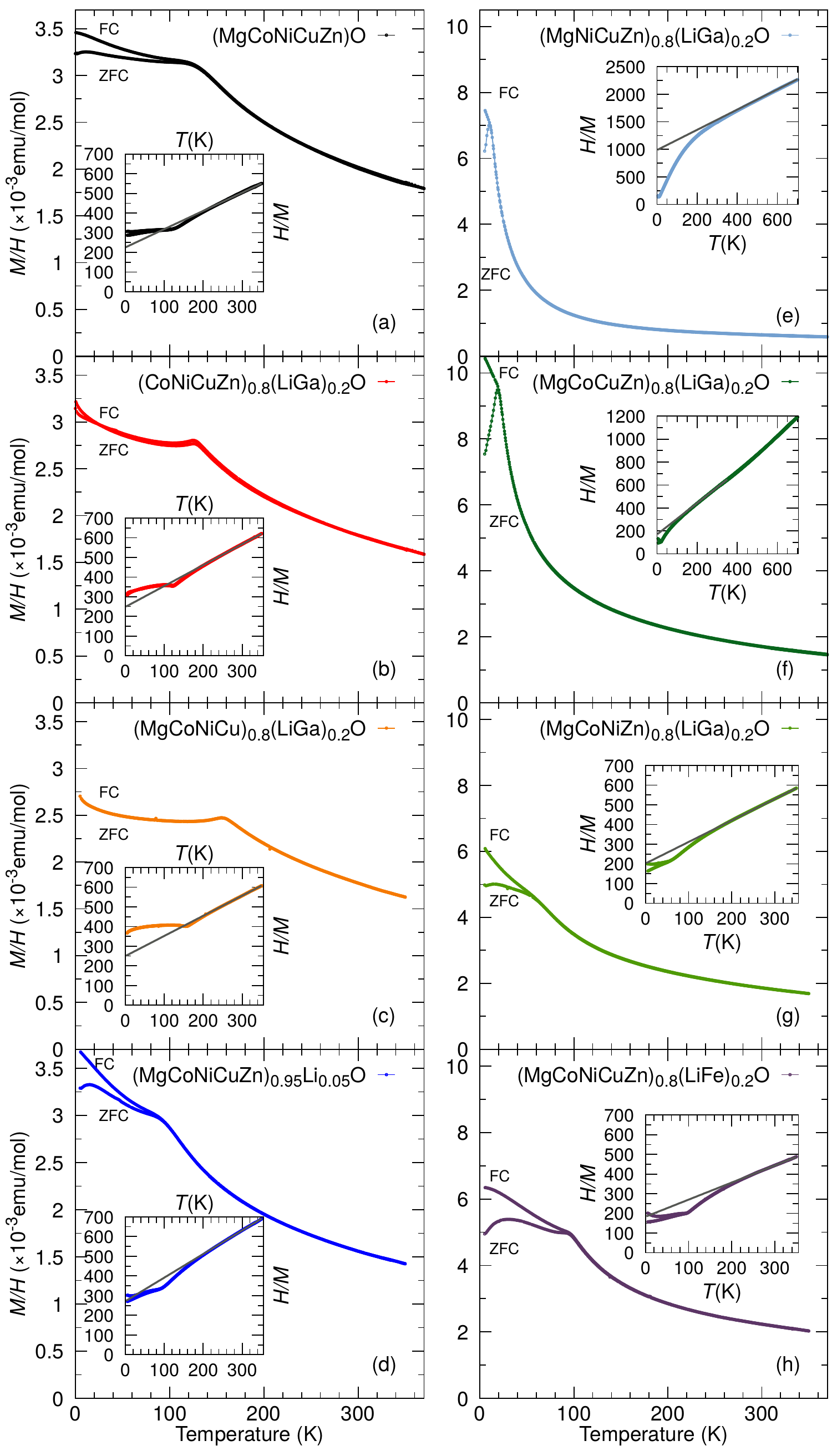}
		\setlength{\abovecaptionskip}{5pt}
		\setlength{\belowcaptionskip}{-14pt}
	\caption{\label{fig:allMT_CW} (color online) Magnetic susceptibility (\textit{M/H}) of the HEOx listed in Table~\ref{tab:series} from (a) to (h) respectively. Lower curves represents measurements of ZFC and upper curves represent measurements of FC. Inset shows the inverse of the magnetic susceptibility and a fitting to the Curie Weiss model $\chi = \frac{C}{T - \Theta} $. }
\end{figure}

\begin{figure}[t]
	\includegraphics[width=8cm]{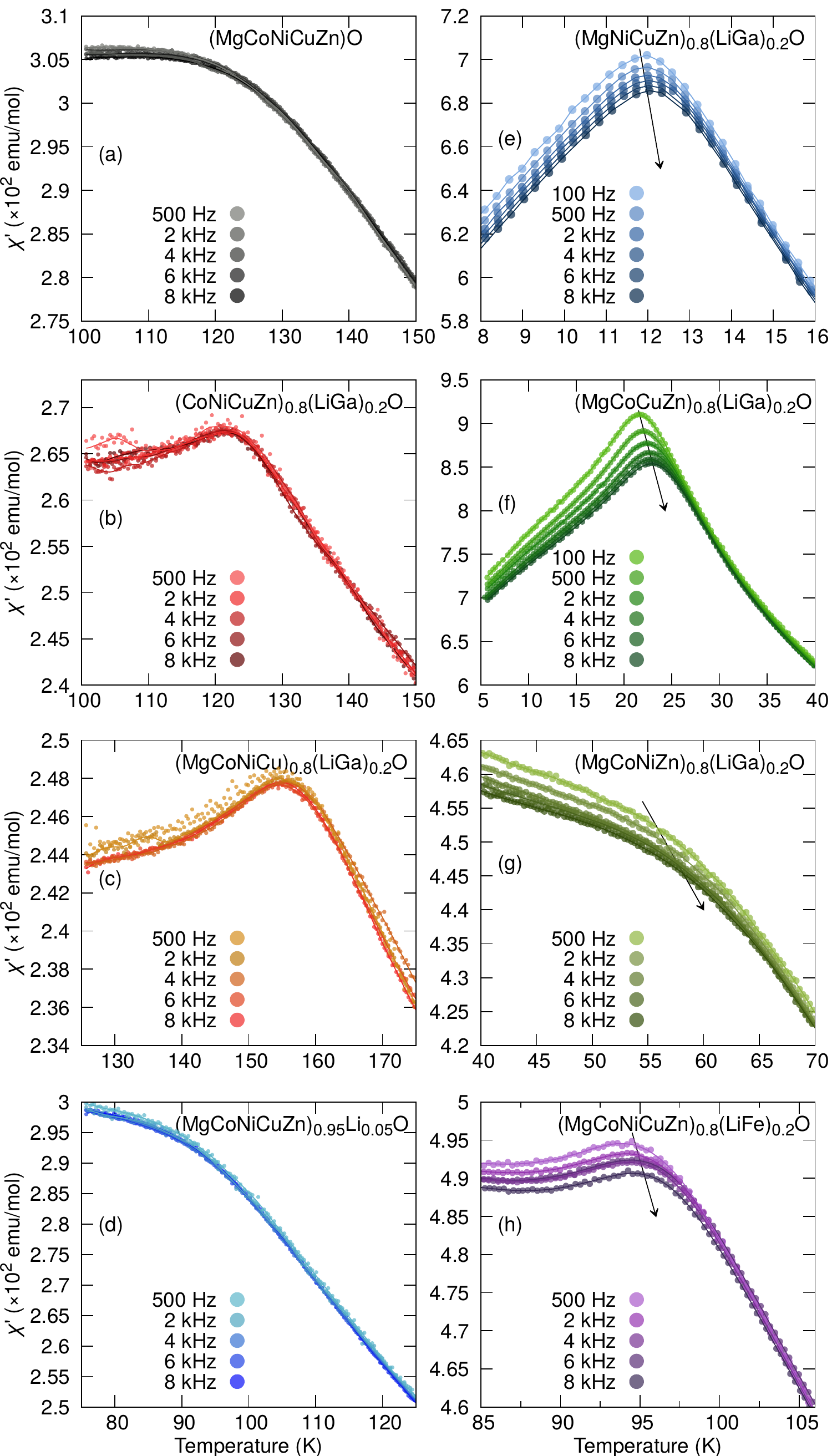}
			\setlength{\abovecaptionskip}{5pt}
			\setlength{\belowcaptionskip}{-14pt}
	\caption{\label{fig:AC_HeOx_all} (color online) Real part of AC susceptibility ($\chi '$) at frequencies between 100 Hz and 8 kHz  of the HEOx listed in Table~\ref{tab:series} from (a) to (h) respectively. The arrows indicate the shift of $T_\text{F}$.}
\end{figure}
%

We have synthesized the series of rocksalt-type HEOx as listed in Table~\ref{tab:series}. All samples were synthesized from binary oxides and carbonates, MgO, Co$_3$O$_4$, NiO, CuO, ZnO, Li$_2$CO$_3$, Fe$_2$O$_3$ and Ga$_2$O$_3$. These starting chemicals were mixed in stoichiometric amounts and then pressed into pellets, which were heated for 12 h at $1000^\circ$~C in air. Finally, they were quenched to room temperature.

\begin{table}[t]
	\caption{\label{tab:series}
		List of high-entropy oxides studied in this paper. All compounds crystallize in rocksalt structure $(Fm\bar{3}m)$.}
	\begin{ruledtabular}
		\begin{center} \small\addtolength{\tabcolsep}{-1pt}
			\scriptsize
			\begin{tabular}{llc}
				Compound (abbreviated)&Chemical formula& cell par. (\AA) \\ \hline
				(MgCoNiCuZn)O               & Mg$_{0.2}$Co$_{0.2}$Ni$_{0.2}$Cu$_{0.2}$Zn$_{0.2}$O                				& 4.236                        \\
				(CoNiCuZn)$_{0.8}$(LiGa)$_{0.2}$O     & Co$_{0.2}$Ni$_{0.2}$Cu$_{0.2}$Zn$_{0.2}$Li$_{0.1}$Ga$_{0.1}$O           & 4.219                        \\
				(MgCoNiCu)$_{0.8}$(LiGa)$_{0.2}$O     & Mg$_{0.2}$Co$_{0.2}$Ni$_{0.2}$Cu$_{0.2}$Li$_{0.1}$Ga$_{0.1}$O           & 4.205                        \\
				(MgCoNiCuZn)$_{0.95}$Li$_{0.05}$O     & Mg$_{0.19}$Co$_{0.19}$Ni$_{0.19}$Cu$_{0.19}$Zn$_{0.19}$Li$_{0.05}$O     & 4.219                        \\
				(MgNiCuZn)$_{0.8}$(LiGa)$_{0.2}$O     & Mg$_{0.2}$Ni$_{0.2}$Cu$_{0.2}$Zn$_{0.2}$Li$_{0.1}$Ga$_{0.1}$O           & 4.214                        \\
				(MgCoCuZn)$_{0.8}$(LiGa)$_{0.2}$O     & Mg$_{0.2}$Co$_{0.2}$Cu$_{0.2}$Zn$_{0.2}$Li$_{0.1}$Ga$_{0.1}$O           & 4.230                         \\
				(MgCoNiZn)$_{0.8}$(LiGa)$_{0.2}$O     & Mg$_{0.2}$Co$_{0.2}$Ni$_{0.2}$Zn$_{0.2}$Li$_{0.1}$Ga$_{0.1}$O           & 4.212                        \\
				(MgCoNiCuZn)$_{0.8}$(LiFe)$_{0.2}$O   & Mg$_{0.16}$Co$_{0.16}$Ni$_{0.16}$Cu$_{0.16}$Zn$_{0.16}$Li$_{0.10}$Fe$_{0.10}$O & 4.219                       
			\end{tabular}
		\end{center}
	\end{ruledtabular}
\vspace{-25pt}
\end{table}

X-ray diffraction measurements were performed with a Panalytical X-Pert instrument equipped with incident Ge monochromator and a copper tube (K$_{\alpha 1}$) and X’celerator detector. Magnetization measurements were conducted by using Magnetic Properties Measurement System (MPMS, Quantum Design) from 5 K to 350~K. The measurements up to 700~K were performed in MPMS3 (Quantum Design) with an oven option. Heat capacity measurements were done with a relaxation method below room temperature down to 5 K by using Physical Properties Measurement System (PPMS, Quantum Design).

The powder neutron diffraction data were collected in the instrument E6 at the BER II reactor of the Helmholtz-Zentrum Berlin für Materialien und Energie. We used a pyrolytic graphite (PG) monochromator selecting the neutron wavelength $\lambda$ = 2.42 \AA. The diffraction patterns were obtained between 1.7 and 150~K in the scattering angle 2$\theta$ ranging from 4.5 to 135.7$^\circ$. Rietveld refinements of the diffraction data were carried out with the program Fullprof  \cite{rodriguez-carvajal_recent_1993}. In the refinement of magnetic structure we used the magnetic form factor of Cu$^{2+}$  \cite{brown_magnetic_1995}.

The X-ray diffraction patterns shown in Fig.~\ref{fig:XRD} indicate that the obtained samples are almost single phase of rocksalt-type oxides. Previous X-ray photoelectron spectroscopy \cite{berardan_colossal_2016},  and electron paramagnetic resonance (EPR)\cite{berardan_controlled_2017} measurements showed that the oxidation states of ions are Mg$^{2+}$, Ni$^{2+}$, Cu$^{2+}$, Zn$^{2+}$, Li$^{+}$, Ga$^{3+}$, Fe$^{3+}$ and Co$^{2+}$ (for undoped samples). In our current study the Co$^{2+}$ ion was found to be in a high-spin state by EPR, which was also suggested by the electronic structure calculations. \cite{rak_charge_2016}. Thus, the electronic configurations of cations that have a total spin different from zero are $3d^5$ $(S = 3/2)$ for Fe$^{3+}$, $3d^7$ ($S = 3/2$ and $l = 1$) for Co$^{2+}$, $3d^8$ $(S = 1)$ for Ni$^{2+}$, and $3d^9$ $(S = 1/2)$ for Cu$^{2+}$, respectively.

Magnetic susceptibility $\chi(T)$ of (MgCoNiCuZn)O measured at 1~kOe, shown in Fig.~\ref{fig:allMT_CW}(a), displays a Curie-Weiss behavior on cooling from room temperature, and a cusp was seen at around $T_\text{mag} \sim  120$~K. The Curie-Weiss fit at high temperatures shown in the inset of Fig.~\ref{fig:allMT_CW}(a) yields the effective moment $\mu_\text{eff} = 2.95 \mu_B$/f.u. and Curie-Weiss temperature $\theta_{\text{CW}} = -245$~K. The magnitude of the effective moment can be accounted for by the presence of 20\% of Ni$^{2+}$, Cu$^{2+}$, and Co$^{2+}$ ions. For Cu$^{2+}$ ions, the local non-cubic distortion in CuO$_6$ octahedra, discerned by the X-ray diffraction and EPR studies \cite{berardan_controlled_2017}, likely lifts the degeneracy of $e_g$ orbitals and produces spin-only $S = 1/2$ moment ($\mu_\text{eff}$ = 1.73 $\mu_B$/Cu). In case of Co$^{2+}$ ions, such distortion was not identified and the cubic environment should be retained. This gives rise to unquenched orbital angular momentum and the resultant spin-orbital-entangled moment in the high-spin state of Co$^{2+}$ $(3d^7)$. The size of magnetic moment for this case depends on the details of local coordinate like the magnitude of crystal field \cite{kanamori_j._1957}. We use the value of effective moment observed in CoO ($\mu_\text{eff} = 4.94 \mu_B/\text{Co}$  \cite{kanamori_j._1957}). Ni$^{2+}$ ion ($3d^8$) does not have orbital degeneracy and gives $S = 1$ moment ($\mu_\text{eff} = 2.83 \mu_B/\text{Ni}$). Thus, the expected effective moment $\mu_{\text{eff.calc.}}$ is, $\mu_\text{eff.calc.}= [0.2\times(1.73)^2 + 0.2\times(2.83)^2 + 0.2\times(4.94)^2]^{1/2} \mu_B \sim 2.66 \mu_B$ which is close to the experimentally obtained value.

Below $T_\text{mag}$,  $\chi(T)$ shows a small bifurcation between zero-field-cooled (ZFC) and field-cooled (FC) curves. On the other hand, the real part of AC magnetic susceptibility $\chi’(T)$ displays a broad peak around $T_\text{mag}$ without any frequency dependence as shown in Fig.~\ref{fig:AC_HeOx_all}(a). This suggests the ground state of (MgCoNiCuZn)O is an antiferromagnetically ordered state rather than a glassy state, despite the large chemical disorder, and the small bifurcation in $\chi(T)$ that likely originates from a small amount of impurities or orphan spins near defects.

\begin{figure}[t]
	\onecolumngrid
	\includegraphics[width=8.8cm]{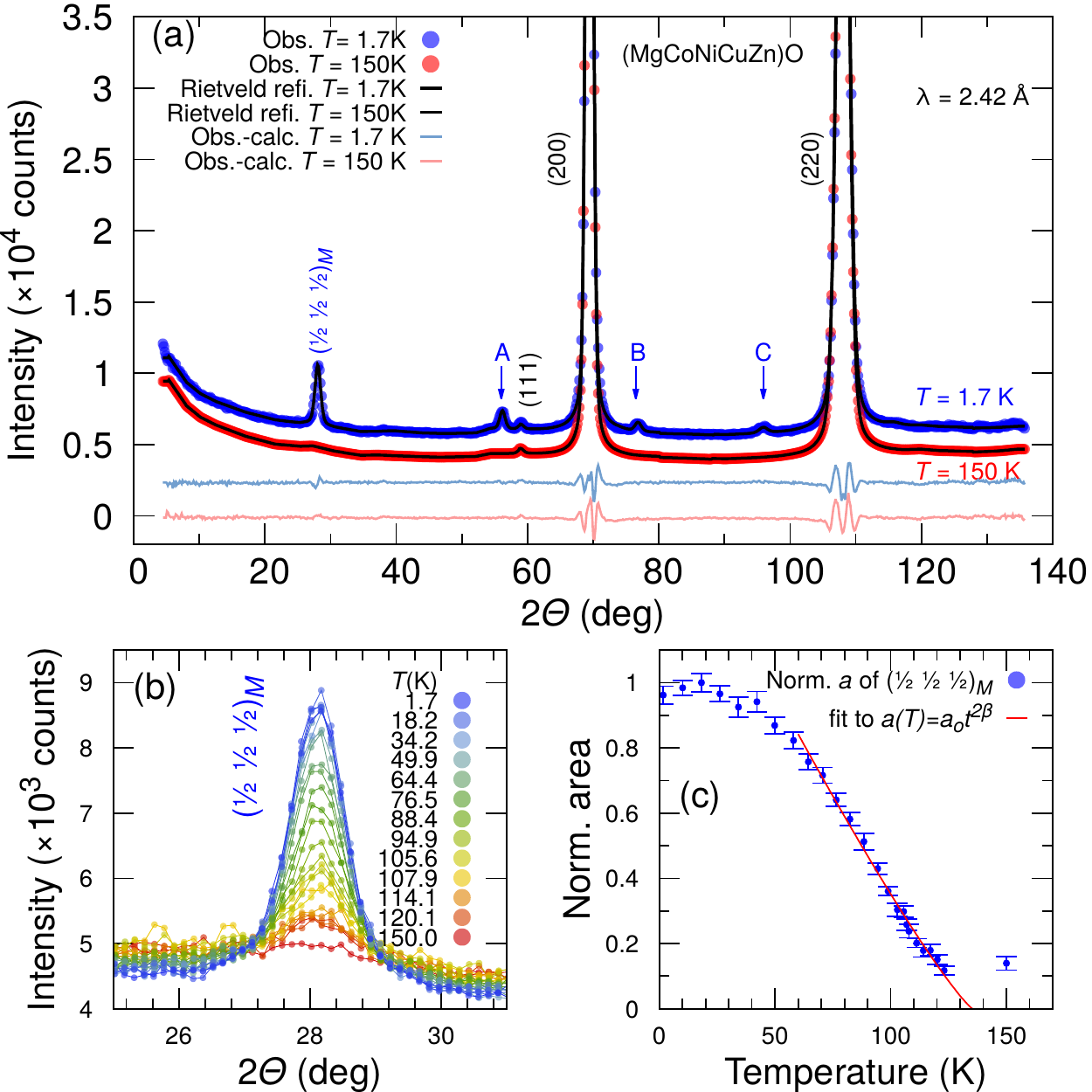}
		\setlength{\abovecaptionskip}{5pt}
		\setlength{\belowcaptionskip}{-14pt}
	\caption{\label{fig:neutrons} (color online) Neutron diffraction of \heoxone . Panel (a) shows the patterns of neutrons diffraction at $T= 1.7$~K and $T= 150$~K). Fitting of Rietveld refinements are drawn with black lines. The difference between the fitting and experiment is drawn at the bottom. The obtained $R_F$ is $R_F = 0.0135$ and  $R_F = 0.0129$ for 150 and 1.7~K respectively where $R_F = \frac{\sum ||F_{\text{obs}}| - |F_{\text{calc}}|| } {\sum|F_{\text{obs}}|} $ . For the magnetic moment $R_M = 0.0503 $ where  		$R_M = \frac{\sum ||I_{\text{obs}}| - |I_{\text{calc}}|| } {\sum|I_{\text{obs}}|} $. Indexes of nuclear and magnetic diffraction are indicated in black and blue, respectively. The blue characters indicate the reflections generated by $\pmb{q} = \left( \frac{1}{2} \frac{1}{2} \frac{1}{2}\right)_M $. A indicates the magnetic reflections at $2\theta =  56.2$~deg $( \bar{1} \bar{1} 1)+\pmb{q}$,  $(  1 \bar{1} \bar{1})+\pmb{q}$ and  $(  0 \bar{2} 0)+\pmb{q}$ . B indicates the reflections at $2\theta = 76.9$~deg $(  1  1 \bar{1})+\pmb{q}$, $( \bar{1}  1  1)+\pmb{q}$ and $(  1 \bar{1}  1)+\pmb{q}$. Finally, C indicates reflections at $2\theta = 95.9$~deg   $(  0  0  2)+\pmb{q}$, $(  0  2  0)+\pmb{q}$,  $(  2  0  0)+\pmb{q}$ and  $(  1  1  1)+\pmb{q}$.  Panel (b) shows the evolution of the magnetic peak $\left( \frac{1}{2} \frac{1}{2} \frac{1}{2}\right)_M $ depending on the temperature. Panel (c) shows the area below the magnetic diffraction $\left( \frac{1}{2} \frac{1}{2} \frac{1}{2}\right)_M $ at different temperatures and a fitting to $a(T)=a_o t^{2\beta}$.}
\end{figure}

The presence of long-range magnetic order was corroborated by neutron diffraction measurements. Fig.~\ref{fig:neutrons}(a) shows the powder diffraction patterns above and below $T_\text{mag}$. At $T = 150~\text{K} > T_\text{mag}$, the peaks of diffraction pattern can be indexed with the nuclear reflections of rocksalt-type structure ($Fm\bar{3}m$). At $T < T_\text{mag}$, the appearance of new peaks was clearly observed. The strongest one at $2\theta \sim 28.8^\circ$ can be indexed with $(\frac{1}{2} \frac{1}{2} \frac{1}{2})$ for the rocksalt-type cubic cell. The other peaks located at 57.0, 77.6, and $96.7^\circ$ were explained by the propagation vector $\pmb{q} = (\frac{1}{2} \frac{1}{2} \frac{1}{2})$ with respect to the reflections such as $(1\bar{1}\bar{1})$, $(11\bar{1})$, and (111), (200), respectively (see caption of Fig.~\ref{fig:neutrons}). Figure~\ref{fig:neutrons}(b) depicts the growth of $(\frac{1}{2} \frac{1}{2} \frac{1}{2})$ reflection below $T_\text{mag} = 120$~K. The integrated intensity of $(\frac{1}{2} \frac{1}{2} \frac{1}{2})$ reflection as a function of temperature shows an order-parameter behavior below $T_\text{mag}$, suggesting a development of long-range magnetic order [Fig.~\ref{fig:neutrons}(c)]. The power-law fit, $a= a_ot^{2\beta}$ where $t=|T_\text{N}-T|/T_\text{N}$   \cite{payne_study_1996,payne_neutron_1997},  yields the Neel temperature $T_\text{N} = 135(4)$~K. The diffraction pattern at 1.7~K in Fig. ~\ref{fig:neutrons}(a) was refined by assuming a magnetic structure similar to that of NiO where ferromagnetically-coupled (111) planes are stacked antiferromagnetically along the $\langle111\rangle$ direction \cite{roth_magnetic_1958}. The refinement of the magnetic structure converges satisfactorily, yielding the ordered moment of $0.83(1) \mu_B /\text{cation}$ lying in the (111) plane [see the caption of the Fig.~\ref{fig:neutrons}(a) for the result of refinement].

\begin{figure}[t]
	\onecolumngrid
	\setlength{\abovecaptionskip}{5pt}	
	\setlength{\belowcaptionskip}{-14pt}
	\includegraphics[width=5.5cm]{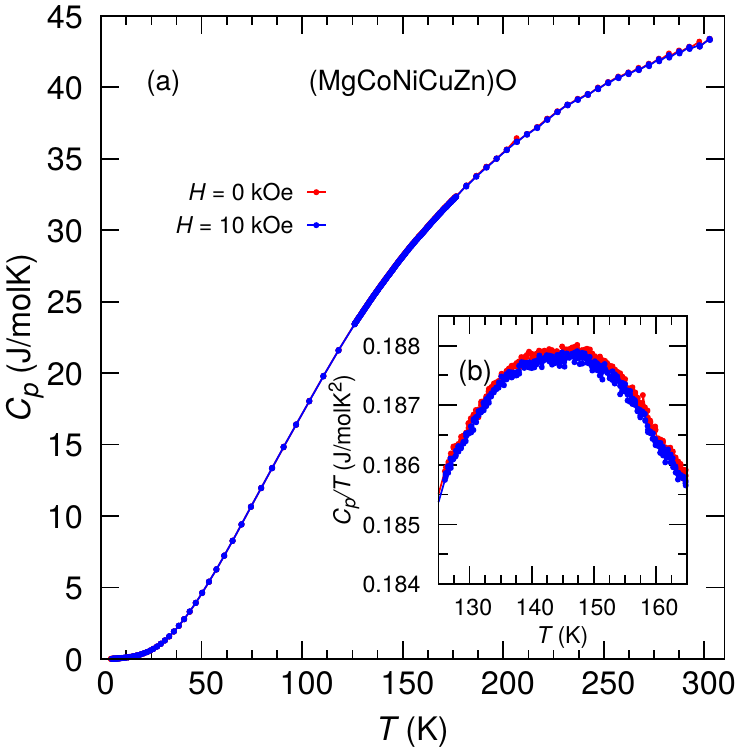}
	\caption{\label{fig:heatc} (color online) Heat capacity of \heoxone . Main panel (a) shows the temperature dependence of the heat capacity ($C_p$ vs $T$) from 2 to 300~K at $H=0$ kOe (red) and $H=10$ kOe (blue). The inset (b) shows $C_p/T$ vs $T$ from 125~K to 165~K.}
\end{figure}

Although the long-range magnetic order of (MgCoNiCuZn)O was evidenced in the neutron diffraction, no anomaly was seen in heat capacity $C_p(T)$ at $T_\text{N}$ as shown in Fig.~\ref{fig:heatc}. This fact suggests the strong fluctuation of magnetic moments. That is, the short-range magnetic correlations develop well above $T_\text{N}$, but the disorder and/or magnetic frustration suppress a formation of long-range magnetic order. The presence of frustration is expected for the FCC lattice of magnetic ions in a rocksalt structure. Indeed, the ratio of Curie-Weiss temperature and magnetic transition temperature, $f = |\theta_\text{CW}|/T_\text{N} \sim 2$, points to a moderate frustration.

The canonical HEOx, (MgCoNiCuZn)O, was therefore found out to exhibit long-range antiferromagnetic order despite the presence of strong disordering of cations. The variation of rocksalt HEOx can be obtained by replacing one of cations with, or by adding, the 1:1 mixture of monovalent and trivalent cations such as Li$^{+}$-Ga$^{3+}$ or Li$^+$-Fe$^{3+}$ as shown in Fig.~\ref{fig:XRD} and listed in Table~\ref{tab:series}. $\chi(T)$ of these HEOx are shown in Fig.~\ref{fig:allMT_CW}(b) to (f). The HEOx containing three magnetic cations shown in the left column [(b) to (d)] display a cusp in $\chi(T)$ at $T_\text{mag} \sim 100-150$~K as in (MgCoNiCuO)O, suggesting the appearance of long-range antiferromagnetic order. $\chi’(T)$ of these HEOx [Fig.~\ref{fig:AC_HeOx_all}(b)-(d)] show no appreciable frequency dependence near the cusp temperature, consistent with the long-range magnetic order. The results of Curie-Weiss fit and magnetic transition temperatures are listed in Table~\ref{tab:properties}. The magnitude of effective moments agree with the calculated value as in (MgCoNiCuZn)O.

On the contrary, the HEOx with only two magnetic ions [Figs.~\ref{fig:allMT_CW}(e)-(g)] display a distinct behavior. $\chi(T)$ shows a sizable bifurcation between ZFC and FC curves at low temperatures below $T\sim25$~K (see Table~\ref{tab:properties}); $\chi(T)$ of the ZFC curve shows a cusp at $T_\text{mag}$, while that of FC process increases monotonically on cooling. $\chi '(T)$ of those HEOx shows a peak at the temperature where the bifurcation of $\chi(T)$ is observed. The peak temperature slightly shifts to a higher temperature by increasing the frequency of AC magnetic fields [see Fig.~\ref{fig:AC_HeOx_all}(e)-(g)]. These suggests a glassy state in these HEOx in contrast to the ones with three magnetic ions. The appearance of glass-freezing can be attributed to the disconnected magnetic exchange paths because of the reduced content of magnetic ions. The isolated magnetic domains likely give rise to a magnetic cluster-glass behavior.

\begin{table}[t]
	\caption{\label{tab:properties}
		Summary of magnetic properties of rocksalt-type HEOx. AF stands for  antiferromagnetic order. $\mu_{\text{eff. calc.}}$ represents the calculated effective moment by assuming the effective moment of each magnetic ions as $5.92 \mu_B$ for Fe$^{3+}$, $4.94 \mu_B$ for Co$^{2+}$, $2.83 \mu_B$ for Ni$^{2+}$ and $1.73 \mu_B$ for Cu$^{2+}$, respectively.}
	\begin{ruledtabular}
		\begin{center} \small\addtolength{\tabcolsep}{-1pt}
			\scriptsize
			\begin{tabular}{lccccc}
				Compound                  & \shortstack{$T_\text{mag}$\\(K)} & \shortstack{Ground\\state} &\shortstack{$\mu_{\text{eff}}$ \\($\mu_B$/f.u.)} &  \shortstack{$\mu_{\text{eff calc}}$ \\($\mu_B$/f.u.)} & \shortstack{$\theta_{\text{CW}}$\\(K)} \\ \hline
				(MgCoNiCuZn)O             & 120      & AF           & 2.95           & 2.66               & -245   \\
				(CoNiCuZn)$_{0.8}$(LiGa)$_{0.2}$O   & 125      & AF           & 2.76           & 2.66               & -242   \\
				(MgCoNiCu)$_{0.8}$(LiGa)$_{0.2}$O   & 155      & AF           & 2.79           & 2.66               & -244   \\
				(MgCoNiCuZn)$_{0.95}$Li$_{0.05}$O   & 95       & AF           & 2.58           & 2.59               & -226   \\
				(MgNiCuZn)$_{0.8}$(LiGa)$_{0.2}$O   & 10       & Spin-glass   & 1.84           & 1.48               & -335   \\
				(MgCoCuZn)$_{0.8}$(LiGa)$_{0.2}$O   & 20       & Spin-glass   & 2.42           & 2.34               & -124   \\
				(MgCoNiZn)$_{0.8}$(LiGa)$_{0.2}$O   & 60       & Spin-glass   & 2.7            & 2.55               & -185   \\
				(MgCoNiCuZn)$_{0.8}$(LiFe)$_{0.2}$O & 100      & ?            & 3.01           & 3.03               & -201  
			\end{tabular}
		\end{center}
	\end{ruledtabular}
\vspace{-25pt}
\end{table}

Fig.~\ref{fig:allMT_CW}(h) shows the magnetic susceptibility of HEOx containing four magnetic ions (Mg$_{0.16}$Co$_{0.16}$Ni$_{0.16}$Cu$_{0.16}$Zn$_{0.16}$)(Li$_{0.10}$Fe$_{0.10}$)O [denoted as (MgCoNiCuZn)$_{0.8}$(LiFe)$_{0.2}$O]. Although $\chi(T)$ shows a similar behavior with the HEOx with three magnetic ions [Fig.~\ref{fig:allMT_CW}(a) or (d)] with a cusp at $T_\text{mag} \sim 100$~K, $\chi’(T)$ of this sample exhibits a small frequency dependence around $T_\text{mag}$. The presence of Fe$^{3+}$, a magnetic ion with different valence state, may provide further disorder in the exchange paths. In order to pin down the magnetic ground state, the neutron diffraction should be required as in (MgCoNiCuZn)O.

The magnetic structure of rocksalt HEOx seems similar to those of binary NiO or CoO. In case of CoO, structural distortion to a monoclinic structure is seen below $T_N$ \cite{jauch_crystallographic_2001}, accompanied by complex magnetic propagation vectors \cite{tomiyasu_magnetic_2004}. The neutron diffraction of (MgCoNiCuZn)O did not show a clear structural distortion at low temperatures, but high-resolution diffraction measurements may be required to clarify the low-temperature crystal structure.

Despite the seemingly similar magnetic properties with binary rocksalt oxides, the magnetic transition temperatures are substantially reduced in the HEOx. This could be understood as a consequence of diluted magnetic ions compared with the binary systems. In addition, HEOx have randomly distributed cations in the FCC lattice, and the different or like-magnetic ions are located next to each other where the magnitude, and even sign, of magnetic exchanges can vary depending on the pairs. This should result in randomly distributed strengths of magnetic exchanges. The presence of magnetic order with a clear cusp in $\chi(T)$ is rather surprising in such a structurally and magnetically randomized system, and is worthy for further investigations. Furthermore, the magnetism of HEOx may provide novel functions as in the recent developments of electric or chemical properties. For example, combined with the colossal dielectric properties, magnetic control of dielectric properties may be realized through magnetodielectric coupling.

In conclusion, the magnetic properties of rocksalt-type HEOx were studied by magnetic and thermodynamic measurements. The prototype material (MgCoNiCuZn)O was found to display a long-range magnetic order in spite of the structural disorder of randomly distributed magnetic ions. This agrees with ref.~\cite{meisenheimer_giant_2017}. The magnetic ground states can be tuned by chemical substitutions. The presence of magnetic order in rocksalt-type HEOx is not only fundamentally intriguing, but we expect that it potentially leads to novel magnetic functions as in the recently-discovered functions of HEOx. We also presume that similar magnetic properties can be realized in HEOx with different structural motifs.

During the preparation of this manuscript, we noticed a related work which gave a similar conclusion to this work \cite{junjie_zhang_aps_2018}.

We thank G. Jackeli and A. Smerald for discussions. This work was partly supported by  the Alexander von Humboldt foundation, Japan Society for the Promotion of Science (JSPS) KAKENHI (No. JP15H05852, JP15K21717, 17H01140), JSPS Core-to-core program “Solid-state chemistry for transition-metal oxides”, and CNRS-Energy program, 2017.

\bibliography{HeOx}

\end{document}